\documentclass[conference]{IEEEtran}
\IEEEoverridecommandlockouts
\usepackage{cite}
\usepackage{xcolor}
\usepackage{tabularx}
\usepackage{amsmath,amssymb,amsfonts}
\usepackage{algorithmic}
\usepackage{graphicx}
\usepackage{dsfont}
\usepackage{booktabs}
\usepackage{siunitx}
\usepackage{multirow}
\usepackage{hyperref}
\usepackage{textcomp}
\usepackage{comment}
\usepackage{xcolor}
\usepackage{booktabs}
\usepackage{multirow}
\usepackage{makecell}   
\def\BibTeX{{\rm B\kern-.05em{\sc i\kern-.025em b}\kern-.08em
    T\kern-.1667em\lower.7ex\hbox{E}\kern-.125emX}}
\begin{document}

\title{Automated Assessment of L2 Speech Rhythm Using Low-Frequency Amplitude Modulations\\
}

\author{\IEEEauthorblockN{João Lima, Lucas Ueda, Paula Costa}
\IEEEauthorblockA{\textit{School of Electrical and Computer Engineering} \\
\textit{Universidade Estadual de Campinas}\\
Campinas, Brazil\\
paulad@unicamp.br}


}

\maketitle

\begin{abstract}
Automated Speaking Assessment of non-native speech must effectively evaluate prosody, including speech rhythm, to align with human perception. However, commonly employed rhythm metrics rely on segmental duration, requiring an additional alignment step, which is error-prone in non-native speech containing disfluencies and mispronunciations. We propose an acoustics-based assessment approach that employs a convolutional neural network to extract rhythm features directly from the speech amplitude envelope, motivated by evidence linking low-frequency modulations to rhythm perception. The proposed models are trained on a proficiency score regression task using the \emph{speechocean762} dataset and compared against duration-based models. Our results show that a model using the amplitude envelope's first derivative achieves the highest correlation with human-assigned scores on the Fluency and Prosody dimensions, producing significantly lower errors than one using segment durations among less fluent speakers. The findings support acoustic envelope features as robust, alignment-free alternatives for L2 rhythm assessment. Code is released publicly.
\end{abstract}


\section{Introduction}
Automated Speaking Assessment (ASA) has become a fundamental component of modern Computer-Assisted Language Learning systems, contributing to increased accessibility in second language (L2) education. Despite substantial advancements in the field, significant challenges remain, since many factors influence listeners’ perception of L2 speech in ways that automatic evaluation methods cannot easily capture from the complex speech signal.

Among these factors, those related to the segmental dimension of speech have been the focus of extensive research, much of which addresses the problem of phone-level pronunciation correctness~\cite{wittPhonelevelPronunciationScoring2000a,huImprovedMispronunciationDetection2015,kimAutomaticPronunciationAssessment2022}. At the same time, the suprasegmental dimension of speech, which encompasses prosodic variability such as intonation and rhythm, is also a relevant indicator of language learners' proficiency. Measures such as speech rate, pause duration, and pitch peak alignment were shown to strongly influence native English listeners’ perception of accentedness in non‑native speech~\cite{trofimovichLEARNINGSECONDLANGUAGE2006}, and prosody-targeted instruction has been found to yield more favorable listener judgments of spontaneous non-native speech than segmental-only training, further underscoring the role of prosody in perceptual significance~\cite{derwingEvidenceFavorBroad1998}. Therefore, ASA systems must be capable of evaluating L2 prosody to produce comprehensive assessments that are aligned with human perception.

The intonation component of prosody is commonly represented by fundamental frequency values (F0). In frame or segment-level evaluations, F0 is extracted and processed for each analysis unit, and modelling is designed to capture intonation contour dynamics from sequences of extracted values~\cite{liMultitaskPretrainingEnhancing, suzukiAutomaticProsodyEvaluation2022, truongAutomaticAssessmentL22018}. Utterance-level analyzes, on the other hand, typically condense this contour into a set of summary statistics such as range, slope, and standard deviation to capture global intonational variability~\cite{blackAutomatedEvaluationNonnative2015,kallioProsodyFluencyFinland2023,dongL2ProsodyAssessment2024}. In contrast, the characterization of speech rhythm, a key prosodic dimension that also contributes to signalling L2 proficiency~\cite{ericksonSpeechRhythmEnglish2013}, remains far less standardized. Traditional approaches rely on duration-based metrics such as vowel-to-consonant ratio, pairwise variability index (PVI), and speech rate, which require pre-extracted phone or syllable boundaries~\cite{laiApplyingRhythmMetrics2013,linImprovingL2English2021a,kyriakopoulosDeepLearningApproach2019a}. However, automatically obtaining these boundaries is particularly error-prone for L2 speech, which frequently contains disfluencies and mispronunciations that deviate from canonical phones used to train alignment models.

Kyriakopoulos et al.~\cite{kyriakopoulosDeepLearningApproach2019a} compared conventional duration-based rhythm metrics, such as PVI, with features learned by a recurrent neural network (RNN) from sequences of vocalic and intervocalic segment durations. The RNN-predicted L2 proficiency scores achieved higher correlations with language test results than those derived from rhythm metrics, indicating that the model captures more generalizable duration dynamics across the utterance. Although promising, this approach still requires forced alignment to obtain segment boundaries, and the authors explicitly identified this step as an important bottleneck, noting that alignment errors can substantially degrade rhythm feature reliability.

Notably, rhythm perception is not governed solely by segmental duration, as listeners also rely on acoustic correlates of prominence. Low-frequency speech modulations captured by amplitude envelopes were shown to reflect syllable-level fluctuations in intensity, playing a central role in the perception of prominence and rhythmic grouping~\cite{cumminsRhythmicConstraintsStress1998a,tilsenLowfrequencyFourierAnalysis2008}. Recent work further argues that a comprehensive account of speech rhythm should incorporate such acoustic prominence cues~\cite{fuchsDurationBasedAcousticSpeech2026}. Importantly, amplitude envelopes can be extracted directly from the speech waveform, eliminating the need for forced alignment and its associated errors on L2 speech for utterance-level ASA. Despite this potential, the amplitude envelope has not yet been exploited as a direct input for automatic L2 rhythm assessment.

In this work, we introduce a perceptually grounded, alignment-free method for L2 rhythm assessment that directly processes the speech amplitude envelope. The experiments are based on a proficiency level regression task performed using human-annotated scores from \emph{speechocean762}~\cite{zhang21x_speechocean}, a publicly available, widely adopted dataset.
The proposed acoustics-based models use a 1D convolutional neural network (CNN) to learn relevant features directly from the extracted envelope. To evaluate this approach, we compare it against a duration-based deep model inspired by Kyriakopoulos et al.~\cite{kyriakopoulosDeepLearningApproach2019a}, which uses forced alignment to obtain segment durations. Both models perform attentional pooling over utterance-level features to produce fixed-size representations, which are mapped to proficiency scores by a fully connected regression head. 

Our main contributions are as follows: (1) we propose an acoustics-based, perceptually grounded deep learning method for automated L2 speech rhythm assessment without requiring forced alignment; (2) we perform a comparative analysis between duration and acoustics-based rhythm features, providing a broader perspective on distinct methods for the characterization of this key prosodic dimension; (3) we make all code resources publicly available, including the implemented models and a stand-alone duration and acoustic rhythm feature extractor, which may serve not only the ASA community but a wider range of speech processing fields\footnote{\url{https://github.com/AI-Unicamp/L2-Speech-Rhythm}}.

\section{Methods}
\label{methods}
\subsection{Task formulation and dataset}
Our experiments are conducted by performing the task of proficiency score regression, a well-established paradigm in the field of ASA. Given an L2 utterance sample represented as a sequence of feature vectors $\mathbf{X} = \{\mathbf{x}_1, \dots, \mathbf{x}_T\}$ and its corresponding human-annotated score $y$ indicating the proficiency level judgment for a given dimension, we train a deep learning model $f$ with parameters $\theta$ to output a scalar prediction $\hat{y} = f(\mathbf{X}; \theta)$. The model is optimized by minimizing a regression loss $\mathcal{L}(\hat{y}, y)$ over the training set so that $\hat{y}$ approximates $y$ accurately.

\begin{figure}[h]
    \centering
    \includegraphics[width=\linewidth]{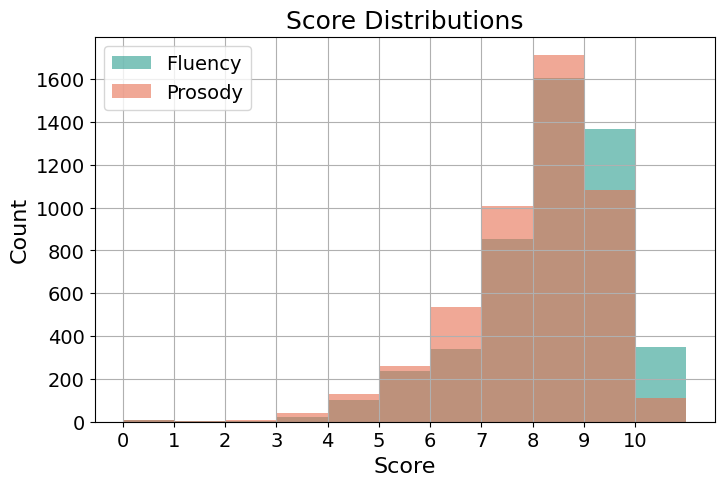}
    \caption{Human-annotated score distributions for the observed proficiency dimensions in the \emph{speechocean762} dataset.}
    \label{fig:scores_histogram}
\end{figure}

To perform the proposed task, we use data from the publicly available \emph{speechocean762} dataset~\cite{zhang21x_speechocean}, which comprises 5000 English speech samples read by 250 learners whose first language (L1) is Mandarin. The mean audio duration is \SI{3.98}{\second}, standard deviation is \SI{1.45}{\second} and each sample was manually annotated by five human experts with scores representing their judgment of the learner's proficiency across multiple dimensions. For speech rhythm assessment, we work with the Fluency and Prosody dimensions, which are annotated based on a rubric observing coherence and occurrence of temporal disfluencies (Fluency), and stability, rhythm, and intonation (Prosody)\footnote{\label{fn:rubric}See Table 1 in~\cite{zhang21x_speechocean} for the full rubric.}. Although the Prosody dimension includes an intonation component not observed in our work, the proposed experiments provide insights into the degree to which rhythm influences listeners' broader prosodic perception. The available scores are scalars ranging from 1 to 10, with higher indicating more proficient speakers, and the distribution is heavily left-skewed, with most scores clustered between 7 and 10 and relatively few low scores, as can be observed in Figure~\ref{fig:scores_histogram}.

\begin{figure*}[t]
    \centering
    \includegraphics[width=\textwidth]{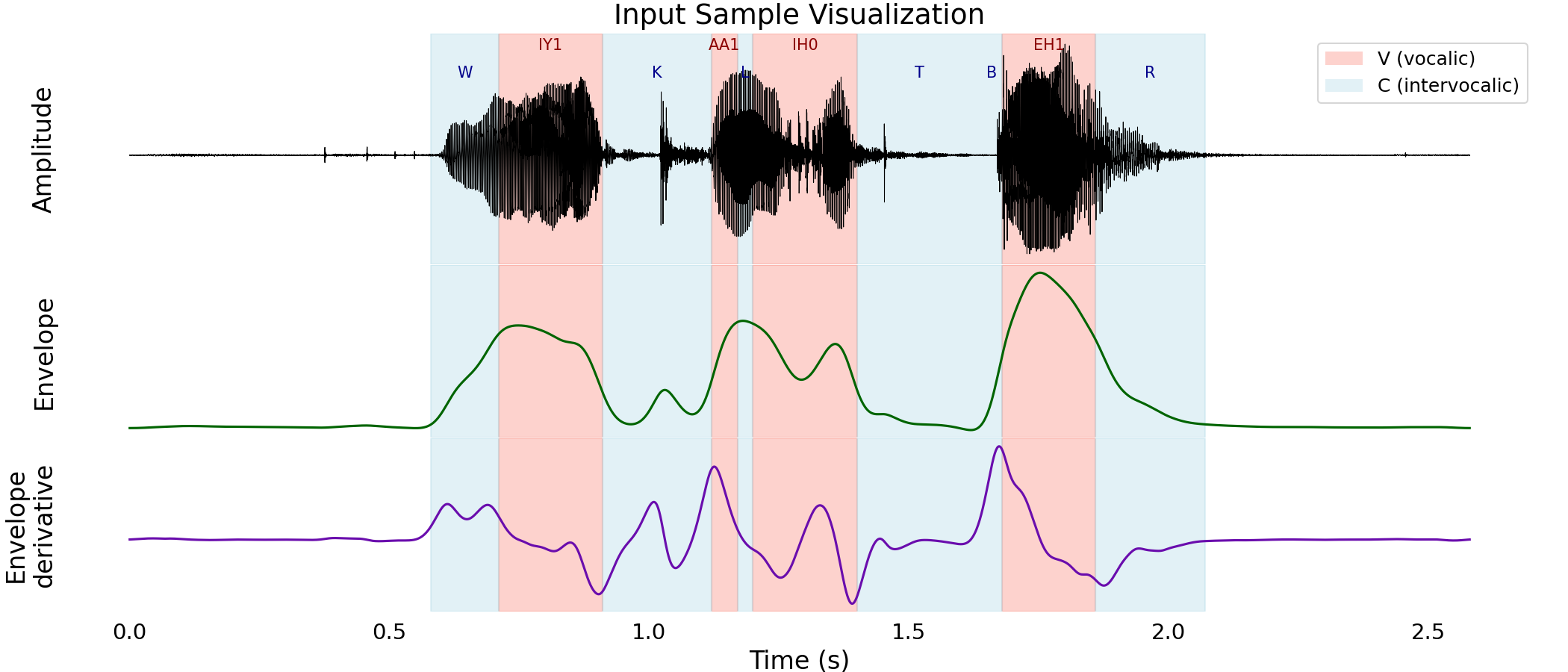}
    \caption{Example audio from the \emph{speechocean762} dataset consisting of a reading of the utterance ``we call it bear". Waveform is shown on the top row. Middle and bottom rows show the extracted amplitude envelope and its first derivative, respectively, which are used as input to the acoustics-based models. Shaded areas in blue and red are vocalic and intervocalic intervals obtained by forced alignment and used as input to the duration-based models. Most intervals have a single phone (sub-interval), except for \{/T/, /B/\}. An alignment error can be observed for the /L/ segment, which incorrectly spans a portion of the preceding vowel instead of the low-energy intervocalic region between AA1 and IH0. Note how peaks in the envelope's derivative are sharper at AA1 and EH1 vowel onset boundaries than the original envelope.}  
    \label{fig:sample_visualization}
\end{figure*}

\subsection{Acoustics-based rhythm features}

Figure~\ref{fig:sample_visualization} shows features extracted from an audio sample using the employed methods.

To obtain acoustics-based rhythm features, we adopt an amplitude envelope extraction method. This choice is motivated by evidence stemming from works across phonetics and neuroscience showing that listeners' perception is closely related to both variability and specific acoustic landmarks in low-frequency regions of the speech signal. Gibbon~\cite{gibbonRhythmsRhythm2023}, as well as Tilsen and Johnson~\cite{tilsenLowfrequencyFourierAnalysis2008}, perform time-frequency and spectral analyzes of amplitude envelopes and observe that they effectively capture hierarchical speech timing, encoding periodicities that range from basic syllables up to long-term discourse, while Cummins and Port~\cite{cumminsRhythmicConstraintsStress1998a} and MacIntyre et al.~\cite{macintyrePushingEnvelopeEvaluating2022} show that specific acoustic landmarks, most notably peaks in the envelope's first derivative, act as accurate proxies for perceptual vowel onsets (p-centers), which are commonly regarded as an anchoring point for temporal coordination in speech. Low-frequency amplitude envelopes have also been shown to relate to human brain responses in the superior temporal gyrus when listening to speech samples~\cite{oganianSpeechEnvelopeLandmark2019}.

We employ the envelope extraction procedure introduced by Schotola~\cite{schotolaUseDemisyllablesAutomatic1984} and build on the publicly available implementation provided by MacIntyre et al.~\cite{macintyrePushingEnvelopeEvaluating2022}, integrating it into our deep learning pipeline. The procedure consists of a psychoacoustically motivated method that simulates key properties of human audition. The speech signal is first decomposed into 22 critical bands spaced on the Bark scale~\cite{zwicker1980analytical}, which reflects the frequency selectivity of the human cochlea. Each band is filtered, squared for rectification, and smoothed with a \SI{1.3}{\milli\second} time constant, then compressed via a logarithmic and square‑root transformation to approximate loudness perception. A weighted sum of band‑specific loudnesses is then performed with positive weights assigned to bands covering the vowel range and negative weights to higher frequencies. This series is then smoothed with forward-backward passes of a 3‑point triangular filter, removing rapid fluctuations while preserving the slower amplitude modulations.

This method was chosen based on the findings of MacIntyre et al.~\cite{macintyrePushingEnvelopeEvaluating2022}, who showed that it can produce envelopes that capture the temporal structure of syllables and stress in English and Mandarin, outperforming simpler envelope extraction methods such as the Hilbert transform.

We also calculate the derivative of the extracted envelope since Oganian and Chang~\cite{oganianSpeechEnvelopeLandmark2019} found that the envelope's rate of change can more reliably signal vowel onsets. Our conducted experiments include a comparison of these features.

\subsection{Duration-based rhythm features}
\label{dur_feats}
To extract duration-based rhythm features, we first employ the Montreal Forced Aligner (MFA)~\cite{mcauliffe17_montrealForcedAligner} to obtain phone boundaries, followed by a segmentation procedure inspired by Kyriakopoulos et al.~\cite{kyriakopoulosDeepLearningApproach2019a}. Given an utterance's phone sequence and its corresponding time alignments, we assign each phone $p_i$ a binary vocalic label $v(p_i) = \mathds{1}[p_i \in \mathcal{V}]$, where $\mathcal{V}$ is the set of vowel tokens in the ARPABET dictionary. We then traverse the sequence and insert an interval boundary wherever $v(p_i) \neq v(p_{i-1})$. This partitions the utterance into a sequence of alternating vocalic (V) and intervocalic (C) intervals, where silence tokens are also categorized as C. See red and blue shaded areas in Figure~\ref{fig:sample_visualization}  for a visualization of V and C intervals, respectively. Each interval $t$ contains a variable number of phones (sub-intervals), denoted by length $L_t$. In the employed dataset, $94.2\%$ of V and $57.2\%$ of C intervals consist of a single phone, and the observed maximum length is $L_{max} = 5$. In the example shown in Figure~\ref{fig:sample_visualization}, all intervals have only one sub-interval except for \{/T/, /B/\}.

For a given interval $t$ containing ordered lists of phone identities and corresponding phone durations, we map each phone identity token into a learnable embedding $\textbf{r} \in \mathbb{R}^{d_{phone}}$ (with $d_{phone}=16$) and concatenate it with the phone’s duration  to compose sub-interval vectors. We experiment with using both raw and z-score normalized phone durations. Z-score normalization is motivated by the fact that phones may naturally differ in duration due to differences in the articulatory movements that produce each type of sound, making deviations from the expected value a more direct measure of prominence. The duration z-score for a phone of identity $p$ with raw duration $d$ is calculated as:
\begin{equation}
z = \frac{d - \mu_p}{\sigma_p}
\end{equation}
where $\mu_p$ and $\sigma_p$ are the mean and standard deviation of raw durations for that specific phone identity token across all observations in the training set. 

The final interval representation $\mathbf{x}_t$ is composed as the concatenation of sub-interval vectors, zero-padded to the maximum interval length $L_{max}$. The sum of all phone durations in the interval is also concatenated as an additional dimension. The resulting vector has dimension $L_{max}(d_{phone}+1) + 1$. 

\subsection{Model architecture}
\label{architecture}
\subsubsection{Acoustics-based models}

A pre-extracted speech envelope $\mathbf{x}_e \in \mathbb{R}^{1 \times T_{\text{in}}}$, where $T_{\text{in}}$ is the input signal length, is first processed by a three-layer 1D CNN with progressive temporal strides of  (3,4,5), yielding a downsampled feature map $\mathbf{M} \in \mathbb{R}^{n_{\text{ch}} \times T_{\text{out}}}$, where $T_{\text{out}} < T_{\text{in}}$ is the downsampled sequence length and $n_{\text{ch}}$ is the number of output channels.

The feature map is then fed into a bidirectional long short-term memory network (BiLSTM) to capture dependencies across the utterance. For each downsampled timestep, the forward and backward hidden states are concatenated into a joint state, yielding a variable‑length sequence. We adopt BiLSTMs in both the acoustic and duration-based architectures to maintain consistency with the original duration‑based approach~\cite{kyriakopoulosDeepLearningApproach2019a}. 

The resulting variable-length sequence is aggregated into a fixed‑size utterance representation via attentive pooling. A learnable linear layer computes a raw score $e_t$ for each timestep. These scores are scaled by a temperature $\tau$, and attention weights are obtained via a sequence‑wise sigmoid normalization:

\begin{equation}
    \alpha_t = \frac{\operatorname{sig}(e_t / \tau)}{\sum_{j} \operatorname{sig}(e_j / \tau) + \epsilon}
\end{equation}
where $\operatorname{sig}(\cdot)$ is the logistic sigmoid function and $\epsilon$ is a small constant for numerical stability. The utterance representation $\mathbf{z}$ is the weighted sum of the hidden states using the learned attention weights, followed by Layer Normalization. Finally, a multi-layer perceptron (MLP) scoring head with a single hidden layer maps $\mathbf{z}$ to a scalar prediction $\hat{y}$ for the proficiency dimension of interest.

Acoustics‑based models trained on envelopes and envelope derivatives are denoted \textit{Env} and \textit{EnvRate}, respectively.

\subsubsection{Duration-based models}

The duration-based models receive two parallel sequences: vocalic $\mathbf{X}^v = (\mathbf{x}^v_1, \mathbf{x}^v_2, \dots)$ and intervocalic $\mathbf{X}^c = (\mathbf{x}^c_1, \mathbf{x}^c_2, \dots)$, where each $\mathbf{x_t}$ is an interval representation constructed as described in Section~\ref{dur_feats}. Both sequences are independently stabilized via Layer Normalization and passed through separate BiLSTM networks. The resulting hidden state sequences are then aggregated into fixed‑size vectors $\mathbf{c}^v$ and $\mathbf{c}^c$ using the same sigmoid‑based attentive pooling described above. These two vectors are concatenated into a joint representation $\mathbf{z} = [\mathbf{c}^v; \mathbf{c}^c]$, which is processed by the MLP scoring head to produce a scalar prediction $\hat{y}$.

Duration‑based models trained with raw and z‑score normalised durations are denoted \textit{Dur} and \textit{DurZ}, respectively.

\section{Experimental setup}
Audio samples are resampled to \SI{16}{\kilo\hertz}, and all rhythm features are extracted prior to training. The extracted amplitude envelopes have a sample rate of \SI{1}{\kilo\hertz}.

All models are trained with the Adam optimizer to minimize Mean Squared Error (MSE) between predicted and ground truth proficiency scores. Fluency and Prosody regression tasks are performed separately. Batch size is fixed at 64, and training is allowed to run for up to 500 epochs, with weight decay and gradient clipping applied to improve stability.

The \emph{speechocean762}  dataset provides predefined train and test splits, however, they are randomly partitioned into equal-sized subsets~\cite{zhang21x_speechocean}, which is suboptimal for deep learning training. We therefore create a new split with a set proportion of 80\% train, 10\% validation, 10\% test, stratified by proficiency score to maintain a similar distribution across all sets.

We monitor the Spearman correlation between ground truth and predicted scores on the validation set during training and set an early stopping trigger if no improvement is observed for 50 epochs. Spearman correlation was adopted as the best model selection criterion due to the skewed distribution of the original scores, and we also report the MSE for the test set in the final evaluation.

Hyperparameter optimization was performed using Optuna over a maximum of 80 trials, with each trial corresponding to an independent training run in the Fluency task using a different hyperparameter configuration. Median pruning was employed to terminate trials whose intermediate validation performance fell below the median performance of previously completed trials at the corresponding training step. The configuration achieving the lowest validation loss was selected for final evaluation on the held out test set. Table~\ref{tab:model_hparams} shows details on the selected hyperparameters.

\begin{table}[h]
\centering
\caption{Hyperparameter configurations used for each model. lr and wd stand for learning rate and weight decay, respectively. The configuration set for the cnn modules is (\textit{output channels, kernel size}). For the bilstm modules, it is (\textit{hidden size, number of layers})}.
\label{tab:model_hparams}

\begin{tabular}{@{}lccccc@{}}
\toprule
Model & LR & WD & CNN & BiLSTM  & Head hidden size\\
\midrule
\textit{Env}
    & $10^{-3}$& $10^{-5}$& (128,7)& (32,2)&128\\

\textit{EnvRate}
    & $10^{-4}$& $10^{-5}$& (128,7)& (128,2)&128\\

\textit{Dur}
    & $5\times10^{-4}$& $10^{-5}$
    & --
    & (64, 1)  &64\\

\textit{DurZ}
    & $5\times10^{-4}$& $10^{-5}$
& --
    & (32,2)&128\\
\bottomrule
\end{tabular}
\end{table}

\section{Results and Discussion}
\begin{figure*}[ht]
    \centering
    \includegraphics[width=0.9\textwidth]{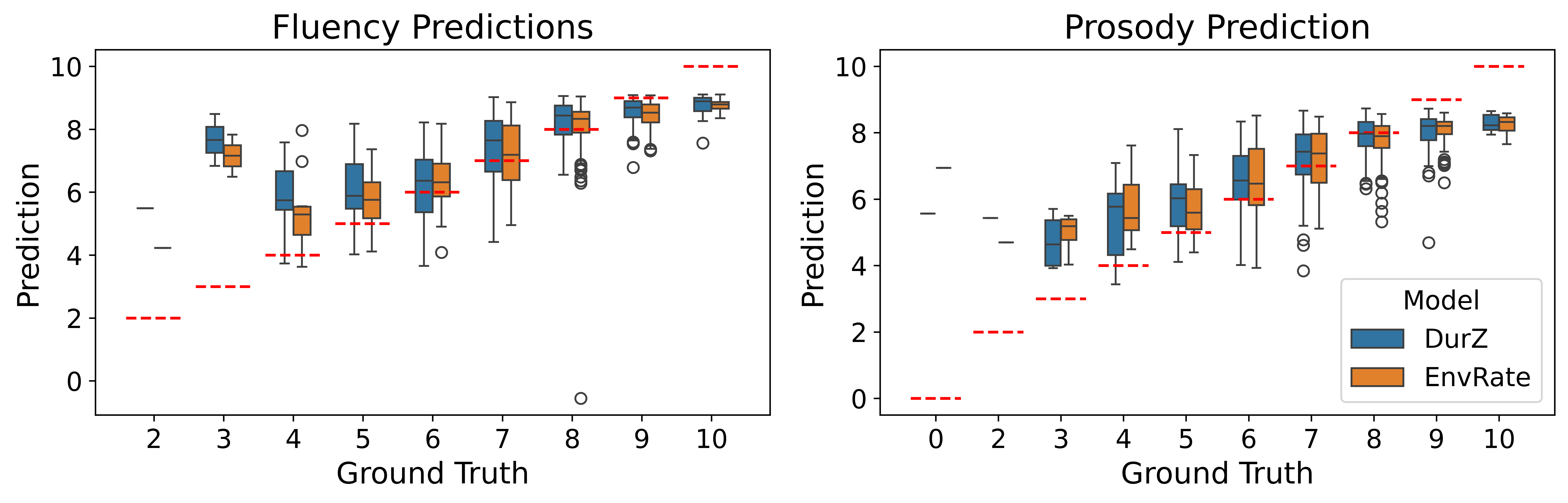}
    \caption{Model predictions by ground truth score for each proficiency dimension in the test set. Red dashed lines indicate human annotated scores.}
    \label{fig:boxplots}
\end{figure*}
Table~\ref{tab:model_results} shows each model's performance at predicting proficiency scores. 

\begin{table}[h]
\centering
\caption{Test set performance across models grouped by proficiency evaluation dimension. Arrows indicate the higher performance direction. Model names appear as defined in Section~\ref{architecture}}.
\label{tab:model_results}

\begin{tabular}{@{}llcc@{}}
\toprule
\textbf{Dimension} & \textbf{Model} & \textbf{Spearman R} ($\uparrow$) & \textbf{MSE} ($\downarrow$) \\
\midrule

\multirow{4}{*}{Fluency}
& \textit{Env}      & 0.65& 0.98\\
\addlinespace[2pt]
& \textit{EnvRate}  & \textbf{0.70}& \textbf{0.97}\\
\addlinespace[2pt]
& \textit{Dur}               & 0.64 & 1.15\\
\addlinespace[2pt]
& \textit{DurZ}             & 0.67& 1.00\\

\midrule

\multirow{4}{*}{Prosody}
& \textit{Env}    & 0.60& \textbf{1.00}\\
\addlinespace[2pt]
& \textit{EnvRate}  & \textbf{0.63}& 1.06\\
\addlinespace[2pt]
& \textit{Dur}              & 0.53& 1.36\\
\addlinespace[2pt]
& \textit{DurZ}             & 0.61& 1.08\\

\bottomrule
\end{tabular}
\end{table}

The \textit{EnvRate} model achieved the best results overall in terms of correlation with human scores for both the Fluency and Prosody dimensions, indicating that the rate of change in the speech envelope is the most powerful feature for the characterization of L2 speech rhythm among those under study. We hypothesize that \textit{EnvRate} achieves better results than \textit{Env} because it provides the model with direct measures of the signal's variability, which is more aligned with listeners' rhythm perception and therefore may capture task-relevant characteristics more easily than instantaneous values. Furthermore, peak locations in the envelope's derivative are more tightly tied to phone (especially vowels) onset locations, as can be observed in Figure ~\ref{fig:sample_visualization}, providing the model with a sharper temporal estimation of relevant acoustic landmarks that drive speech rhythm perception. This is aligned with the findings presented by Oganian and Chang showing that listeners' neural response is synchronized with peaks in the envelope's derivative~\cite{oganianSpeechEnvelopeLandmark2019}.

Between the duration models, the one using z-scores performs better than the one using raw duration values, which is on par with expectations since z-scores may provide the model a more direct measurement of a phone's prominence for signalling L2 proficiency.

Model performance for the Prosody dimension is worse than for Fluency across all experiments. This may be due to the fact that intonation is included under the Prosody human evaluation rubric in the dataset, and thus it influences the ground truth scores in ways that can't be directly modelled by our approach, focused on speech rhythm. However, an inter-dimension comparison of each model's performance reveals that the performance gap between dimensions is narrower for the acoustics-based and larger for the duration-based models, indicating that the former not only surpasses the latter at L2 rhythm categorization, but is also more tightly related to broader prosodic phenomena in L2 speech.

Figure~\ref{fig:boxplots} shows predicted score distributions across ground truth scores in the test set. There is an evident positive correlation between both models' predictions and the targets, as indicated by the Spearman R metrics in Table~\ref{tab:model_results}. However, model performance is severely degraded when predicting low proficiency scores. This may be due to two disparate possible causes or a combination of them. The first is that there are very few low proficiency samples in the dataset, as can be observed in Figure~\ref{fig:scores_histogram}, which makes it naturally harder for the models to learn generalizable features that can reliably score these samples. The second is that eventual irregularities and disfluencies present in low proficiency L2 speech may challenge not only the duration-based models, but also the envelope-based ones.

Although both models exhibit this degradation, Figure \ref{fig:boxplots} suggests that \textit{EnvRate} Fluency predictions are consistently closer to the target values in the $[2, 5]$ region than \textit{DurZ} predictions. To inspect this, we split the test set into Low and Medium-High proficiency subsets and compute, for each subset, the MSE of each model together with a 95\% percentile bootstrap confidence interval. Low proficiency samples are defined as those having a ground truth score $\leq5$ for Fluency and $\leq6$ for Prosody. They are Medium-High otherwise. The threshold values are chosen based on the original rubric annotators used to score the samples~\footref{fn:rubric}. For each proficiency group $g$, we also compute the mean per-sample squared-error difference $\bar{d}_{g}$, defined as:

\begin{equation}
\bar{d}_{g} = \frac{1}{n_g} \sum_{i}
 \bigl(\hat{y}_i^{\text{DurZ}} - y_i\bigr)^2
     - \bigl(\hat{y}_i^{\text{EnvRate}} - y_i\bigr)^2
\end{equation}

Where $y_i$ is the ground truth score, $\hat{y}_i^\cdot$ is the prediction from each model, and $n_g$ is the number of samples in group $g$.

A paired Wilcoxon signed-rank test was performed under the null hypothesis that the per-sample error differences in the group were symmetric about zero. We perform this analysis with only the \textit{EnvRate} and \textit{DurZ} models, as they were shown to be the best performing ones of their respective categories.

Table~\ref{tab:mse_score_groups} reports the MSE for each model across the Low (L-Prof) and Medium-High (MH-Prof) proficiency groups. There is a clear distinction between groups: L-Prof exhibits higher errors for both Fluency and Prosody, reflecting the performance degradation shown in Figure~\ref{fig:boxplots}. The observed differences between models in the L-Prof Fluency region was statistically significant ($p<0.05$, $n=37$), meaning that \textit{EnvRate} produces significantly lower errors in this group. This finding is supported by the mean paired error $\bar{d}_g=1.47$, whose 95\% confidence interval excludes zero.
In the remaining groups, the models perform similarly. \textit{DurZ} shows a lower MSE in the L-Prof Prosody and MH-Prof Fluency groups, while \textit{EnvRate} shows a lower error in the MH-Prof Prosody group. However, none of these differences reach statistical significance, indicating that the models achieve comparable performance in these regions despite the observed point differences.

\begin{table}[t]
\centering
\caption{Mean squared error grouped by low (l-prof) and medium-high (mh-prof) ground truth score groups. Lower is better. bold marks the better model within each group. Asterisks ({\large$\ast$}) indicate pairs with statistically significant differences. $\bar{d}_{g}$ reports the mean paired error difference (DurZ $-$ EnvRate).}
\label{tab:mse_score_groups}
\begin{tabular*}{\linewidth}{@{\extracolsep{\fill}}ll@{\hspace{0.1em}}cc@{}}
\toprule
\textbf{Dimension} & \textbf{Model} & \textbf{L-Prof [95\% CI]} & \textbf{MH-Prof [95\% CI]} \\
\midrule
\multirow[t]{3}{*}{Fluency}
& \textit{EnvRate} & \textbf{2.85} [1.49, 4.54]\rlap{\large$\ast$} & 0.82 [0.60, 1.18] \\
& \textit{DurZ}    & 4.32 [2.58, 6.31]\rlap{\large$\ast$}         & \textbf{0.75} [0.66, 0.85] \\
\cmidrule[0.2pt]{3-4}
& $\bar{d}_{g}$ &  $1.47$ [0.07, 3.00] &  $-0.07$ [$-$0.45, 0.17] \\
\midrule
\multirow{3}{*}{Prosody}
& \textit{EnvRate} & 3.37 [1.72, 5.87]          & \textbf{0.83} [0.73, 0.94] \\
& \textit{DurZ}    & \textbf{3.17} [1.90, 4.81]  & 0.89 [0.76, 1.03] \\
\cmidrule[0.1pt]{3-4}
& $\bar{d}_{g}$ &  $-0.20$ [$-$1.41, 0.85] & $0.05$ [$-$0.08, 0.20] \\
\bottomrule
\end{tabular*}
\end{table}

\section{Conclusions}
In this work, we introduced a perceptually grounded, alignment-free approach to L2 speech rhythm assessment. We showed that envelope-derived features extracted by a CNN offer a strong alternative to segment durations for capturing the rhythmic dimension of L2 proficiency, achieving the highest correlations with human judgments without requiring an extra phone alignment step. A model that used envelope-derived features produced significantly lower errors than one using segment durations when scoring less proficient speakers in the Fluency dimension, suggesting it is more robust to the performance degradation commonly observed for this group, while performing comparably in the other groups.

Our analysis is limited to read English speech samples and Mandarin L1 speakers, therefore future work would benefit from further investigations on how these findings generalize to the more challenging scenario of spontaneous speech with more diverse non-native speakers, as well as to distinct L2s other than English and longer audio samples.

Possible avenues for further investigation also include inspecting the learned attention weights along the input sequences, which could help pinpoint specific regions in the signal relevant for characterizing proficiency, as well as analysing the interaction between the learned acoustic rhythm features and intonation features in a broader assessment scenario.

While this work is concerned with L2 speech, the question of how to model speech rhythm is relevant to a wider range of speech processing areas. Previous works have shown that timing is informative for speech impairment detection~\cite{gogoi25_interspeech} and carries speaker-specific information relevant to applications such as forensic voice comparison and voice anonymization~\cite{tomashenko2025analysis, tomashenko25_interspeech}, so comparing duration-based and acoustic rhythm representations across these domains may prove valuable.

 \section{Acknowledgements}
This study was financed in part by the Coordenação de Aperfeiçoamento de Pessoal de Nível Superior - Brasil (CAPES) - Finance Code 001, and by the São Paulo Research Foundation (FAPESP), Brasil, through process number \#2025/09592-5 and Grant number \#2023/12865-8, Horus project. The experiments were conducted using computational resources provided by the Recod.ai Artificial Intelligence Laboratory at UNICAMP.

\textit{AI-Generated Content Disclosure}: This work was partially developed with assistance from generative AI tools. The Deepseek-V4 model was used to assist with code implementation for the methods described in Section~\ref{methods}. AI tools were also used to polish this manuscript's grammar, clarity, and readability. All AI-generated content was reviewed and validated by the authors. 



\bibliographystyle{IEEEtran}
\bibliography{references}

\end{document}